\documentclass[a4paper,11pt]{article}
\pdfoutput=1 
\usepackage{jheppub} 
\usepackage[T1]{fontenc} 
\usepackage{amsmath,amssymb,epsfig,ulem,color,slashed,wasysym}
\allowdisplaybreaks 

\def \beq{\begin{equation}}
\def \eeq{\end{equation}}
\def \bea{\begin{eqnarray}}
\def \eea{\end{eqnarray}}

\def\bm#1{\mbox{\boldmath$#1$\unboldmath}} 

\title{Constraints on the  trilinear  Higgs  coupling \\from vector boson fusion and associated \\ Higgs production at the LHC}

\author[1]{Wojciech Bizo\'{n},}
\author[2]{Martin Gorbahn,}
\author[1,3]{Ulrich Haisch}
\author[1,3]{and Giulia Zanderighi}

\affiliation[1]{Rudolf Peierls Centre for Theoretical Physics,
   University of Oxford, \\ OX1 3NP Oxford, United Kingdom}

\affiliation[2]{Department of Mathematical Sciences, University of Liverpool, 
\\ L69 7ZL Liverpool, United Kingdom}
    
\affiliation[3]{CERN, Theoretical Physics Department, \\ CH-1211 Geneva 23, Switzerland}

\emailAdd{Wojciech.Bizon@physics.ox.ac.uk}
\emailAdd{Martin.Gorbahn@liverpool.ac.uk}
\emailAdd{Ulrich.Haisch@physics.ox.ac.uk}
\emailAdd{Giulia.Zanderighi@physics.ox.ac.uk}

\abstract{We examine the constraints on  the trilinear Higgs coupling $\lambda$ that originate from associated~($Vh$) and vector boson fusion~(VBF) Higgs production in $pp$ collisions in the context of the Standard~Model effective field theory. The 1-loop contributions to $pp  \to V h$ and $pp \to jj h$ that stem  from insertions of the dimension-6 operator $O_6 = - \lambda \left (H^\dagger H \right )^3$ are calculated and combined with the ${\cal O} (\lambda)$ corrections to the partial decay widths of the Higgs boson. Employing next-to-next-to-leading order QCD predictions, we  analyse the sensitivity of current  and forthcoming measurements of the signal strengths in  $Vh$ and~VBF~Higgs production  to changes in $\lambda$. We show that future~LHC runs  may be able to  probe  modifications of~$\lambda$ with a sensitivity similar to the one that is expected to arise from determinations of double-Higgs production. The sensitivity of differential $Vh$ and VBF~Higgs distributions to  a modified $h^3$ coupling is also studied.}

\preprint{CERN-TH-2016-215, LTH 1104, OUTP-16-23P}

\begin{document} 

\maketitle

\flushbottom

\section{Introduction}
\label{sec:introduction}

Within the Standard Model (SM), the mass and the self-interactions  of the  Higgs field $h$ are parametrised by  the potential 
\beq \label{eq:VSM}
{\cal L}_{\rm SM} \supset -V_{\rm SM} = -\frac{m_h^2}{2} \hspace{0.5mm} h^2 - \lambda v \hspace{0.25mm} h^3 - \frac{\kappa}{4} \hspace{0.5mm} h^4 \,, 
\eeq
where  $v = 246.22 \, {\rm GeV}$ denotes  the Higgs vacuum expectation value  and 
\beq \label{eq:l34}
 \lambda = \kappa = \frac{m_h^2}{2v^2} \,.
\eeq
The LHC measurement  of the Higgs-boson mass  $m_h = 125.09 \, {\rm GeV}$~\cite{ATLASCMS} determined the first term in (\ref{eq:VSM}), but the $h^3$ and $h^4$ couplings, and in particular the SM relation (\ref{eq:l34}) have not been tested. Trying to constrain the Higgs self-couplings and thereby exploring the mechanism of electroweak symmetry breaking (EWSB) is hence an important goal of forthcoming LHC runs  and other future high-energy colliders such as a  hadron-hadron Future Circular Collider  or  a Circular Electron-Positron Collider. 

One way to constrain the coefficients $\lambda$ and $\kappa$ in (\ref{eq:VSM}) consists in measuring double-Higgs and triple-Higgs production.  Since the cross section for $pp \to 3h$ production is of ${\cal O} (0.1 \,{\rm fb})$ at $14 \, {\rm TeV}$ centre-of-mass energy ($\sqrt{s}$) even the high-luminosity option of the LHC~(HL-LHC) will only be able to set very loose bounds on the Higgs quartic.  The prospect to observe double-Higgs production at the HL-LHC is considerably better because at~$14 \, {\rm TeV}$ the $pp \to hh$ production cross section amounts to~${\cal O} (35 \,{\rm fb})$~\cite{Glover:1987nx,Plehn:1996wb,Dawson:1998py,Djouadi:1999rca,deFlorian:2013jea,Grigo:2013rya,Borowka:2016ehy,Borowka:2016ypz}. Measuring double-Higgs production at the HL-LHC however still remains challenging~(see for instance~\cite{Baur:2002qd,Baur:2003gp,Dolan:2012rv,Baglio:2012np,Barr:2013tda,Dolan:2013rja,Papaefstathiou:2012qe,Goertz:2013kp,Maierhofer:2013sha,deLima:2014dta,Englert:2014uqa,Liu:2014rva,Goertz:2014qta,ATL-PHYS-PUB-2014-019,Azatov:2015oxa,Dall'Osso:2015aia,ATL-PHYS-PUB-2015-046,Kling:2016lay}) and as a result even with the full data set of~$3 \, {\rm ab}^{-1}$ only  an ${\cal O} (1)$ determination of the trilinear Higgs coupling seems possible under optimistic assumptions. 

A second possibility consists in studying the effects that a modification of $\lambda$ has at loop level in single-Higgs production. In fact, such indirect probes of the~$h^3$ coupling have been first proposed in the context precision studies of~$e^+ e^- \to hZ$~\cite{McCullough:2013rea,Shen:2015pha} and subsequently extended to observables accessible at hadronic machines such as the LHC~\cite{Gorbahn:2016uoy,Degrassi:2016wml}. For both types of colliders it has been shown that future determination of $\lambda$ via loop effects are complementary to the direct HL-LHC determination through $pp \to hh$, since these probes can provide competitive constraints  under the simplified assumption that new-physics effects dominantly modify the $h^3$ coupling. 

This paper is a sequel to the article \cite{Gorbahn:2016uoy}, in which  two of us have calculated  the ${\cal O} (\lambda)$ corrections to the $gg \to h$ and $h \to \gamma \gamma$  processes that arise at the 2-loop level within the SM effective field theory~(SMEFT). The discussion in the present paper focuses instead  on associated ($Vh$) and vector boson fusion~(VBF) Higgs production.  Specifically,  we compute the 1-loop contributions to the $pp \to Vh$ and $pp \to jj h$ amplitudes that result from insertions of the effective operator~$O_6 = - \lambda \left (H^\dagger H \right )^3$. Combining these contributions with the ${\cal O} (\lambda)$ corrections to the partial decay widths of the Higgs boson, we analyse the sensitivity of present and future~LHC measurements of the $Vh$ and VBF~Higgs processes to shifts in the trilinear Higgs interactions. In order to obtain high-precision predictions for the~$Vh$ and VBF~Higgs cross sections we include~QCD corrections up to next-to-next-to-leading order (NNLO) in our study. We find  that HL-LHC measurements of the~$Vh$ and VBF signal strengths may allow to set bounds on the Wilson coefficient  of $O_6$ that are comparable to the limits that are expected to arise from HL-LHC determinations of $pp \to hh$. By studying differential distributions  it may even be possible to improve the obtained constraints. We present NNLO predictions for the $Vh$ and VBF~Higgs distributions that are most sensitive to the shifts in the trilinear Higgs interactions. Our analysis shows that measurements of the spectra in $Vh$ production provide sensitivity  to the relative sign of the Wilson coefficient of $O_6$. The discriminating power in VBF Higgs production is less pronounced compared to the $Vh$ channels. A similar investigation of the~$Vh$ and~VBF~Higgs processes in an anomalous coupling approach was  presented in~\cite{Degrassi:2016wml}. Whenever indicated we will highlight the similarities and differences between this and our work. 

The article at hand is structured in the following way. In Section~\ref{sec:preliminaries} we introduce the  effective interactions relevant for the computations performed in our paper. The results of our loop  calculations of the~$VVh$ vertex  and the partial decay widths of the Higgs boson are presented in Section~\ref{sec:VVh} and~\ref{sec:widths}, respectively. The computations of the vector boson mediated Higgs cross sections and distributions are described in~Section~\ref{sec:Vh} and~\ref{sec:VBF}. Our numerical analysis  is presented in Section~\ref{sec:numerics}. Both LHC Run I and HL-LHC constraints on the trilinear Higgs coupling are considered.  Section~\ref{sec:conclusions} contains our conclusions. 

\section{Preliminaries}
\label{sec:preliminaries}

Physics beyond the SM~(BSM) can be described in a model-independent fashion by supplementing the SM Lagrangian ${\cal L}_{\rm SM}$ by effective operators $O_k$ of mass dimension six. In our article, we will consider the following Lagrangian
\begin{equation} \label{eq:Lall}
{\cal L}=  {\cal L}_{\rm SM} + \sum_{k=6,H}  \, \frac{\bar c_k}{v^2} \, O_k  \,,
\end{equation}
where 
\begin{equation} \label{eq:operators}
O_6 = - \lambda \, \big ( H^\dagger H \big )^3 \,, \qquad 
O_H = \frac{1}{2}  \, \partial_\mu \big ( H^\dagger H \big ) \, \partial^\mu \big ( H^\dagger H \big ) \,,
\end{equation}
with $\lambda$ defined as in (\ref{eq:l34}) and  $H$ denoting the SM Higgs doublet. The  dimension-6 operators introduced in (\ref{eq:operators}) modify the trilinear Higgs coupling. Upon canonical normalisation of the Higgs kinetic term, one finds 
\beq \label{eq:Ltriple}
{\cal L}  \supset  - \lambda c_3  \hspace{0.25mm} v  \hspace{0.25mm} h^3 = - \lambda \left ( 1 + \bar c_6 - \frac{3 \bar c_H}{2} \right )   \hspace{0.25mm} v  \hspace{0.25mm} h^3 \,,
\eeq
where the Wilson coefficients $\bar c_6$ and $\bar c_H$ as well as the trilinear Higgs coupling $\lambda$ are all understood to be evaluated at the weak scale hereafter denoted by $\mu_w$. 

 It is important to realise  that  the indirect probes of the trilinear Higgs coupling considered in our work measure $c_3$,~i.e.~the coefficient multiplying the interaction term $- \lambda \hspace{0.25mm} v  \hspace{0.25mm} h^3$ in the effective Higgs potential after EWSB. Relating the coefficient $c_3$ to any underlying theory, such as for instance the SMEFT, necessarily involves model assumptions. In the following we will focus our attention on BSM scenarios where the Wilson coefficient $\bar c_6$ represents the only relevant modification of the $h^3$ vertex.  Corrections due to $\bar c_H$ are on the other hand ignored. Such effects will cause a universal shift in all Higgs-boson couplings at tree level and also induce logarithmically-enhanced contributions to the oblique parameters~$S$ and $T$ at the 1-loop level~\cite{Elias-Miro:2013eta}. The Wilson coefficient $\bar c_H$ can therefore be probed by means other than $Vh$ or VBF~Higgs production  that are the focal point of the present work. We also do not consider effects of dimension-8 operators such as $- \lambda \bar c_8/v^4 \, \big ( H^\dagger H \big )^4$.\footnote{The effects of $- \lambda \bar c_8/v^4 \, \big ( H^\dagger H \big )^4$ could be easily incorporated in our analysis by shifting the coefficient~$c_3$ introduced in (\ref{eq:Ltriple}) by $2 \bar c_8$ \cite{Gorbahn:2016uoy}.} Under these model assumptions one obtains the simple relation 
\beq \label{eq:c3toc6bar}
c_3 = 1 + \bar c_6 \,, 
\eeq
which allows one to parameterise modifications of the $h^3$ vertex in terms of the Wilson coefficient $\bar c_6$. In our article we will use this parameterisation, but emphasise that all formulas and results presented in the following sections can  be translated to an anomalous coupling approach by simply replacing $\bar c_6$ with $c_3 -1$. In fact, we have verified that to the perturbative order considered here and in~\cite{Degrassi:2016wml} the calculations of $Vh$ and VBF Higgs production in the SMEFT and the anomalous coupling framework agree exactly if the relation (\ref{eq:c3toc6bar})  is taken into account. 

\section{Corrections to the $\bm{VVh}$ vertex}
\label{sec:VVh}

In the SMEFT there can be  two different types of corrections to the $V V h$ vertex with $V = W,Z$. First, terms that are enhanced by  logarithms of the form $\ln (\Lambda^2/\mu_w^2)$ which are associated to the renormalisation group evolution that  connects the new-physics scale~$\Lambda$  to~$\mu_w$ and second, finite contributions that originate from the corrections to the $VVh$ Green's function with a modified $h^3$ vertex. Since the operator $O_6$ only mixes with itself at the 1-loop level~\cite{Elias-Miro:2013gya,Jenkins:2013zja,Jenkins:2013wua,Alonso:2013hga}, the~$V V h$ vertex does not receive logarithmically-enhanced corrections proportional to $\bar c_6$ at the first non-trivial order in perturbation theory.  
 
\begin{figure}[!t]
\begin{center}
\includegraphics[width=0.75\textwidth]{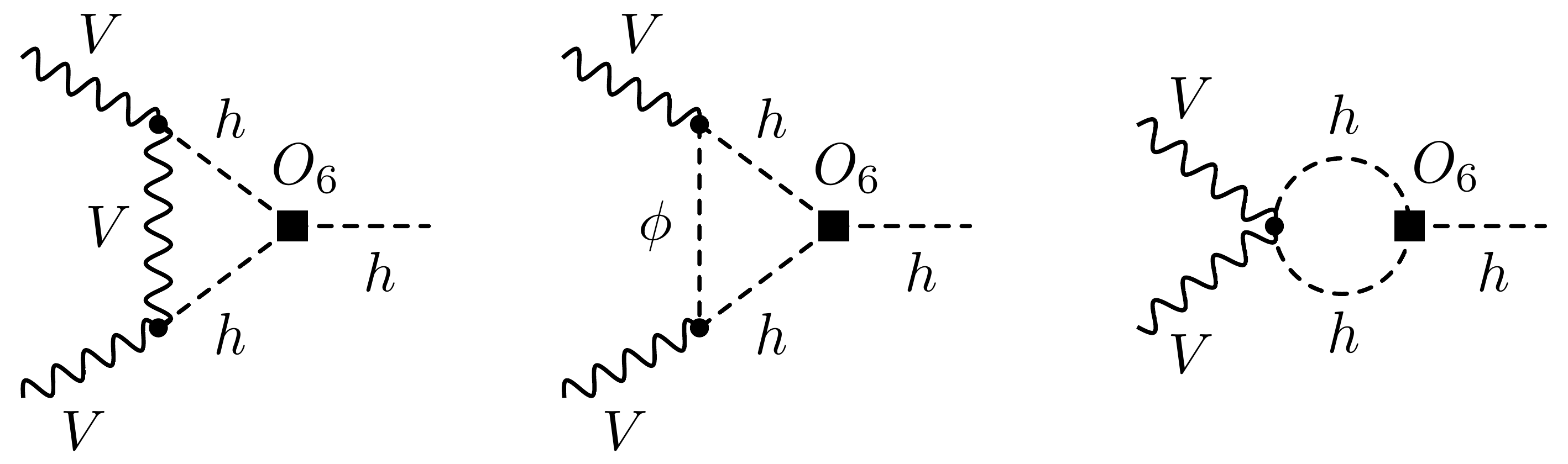}  
\vspace{2mm}
\caption{\label{fig:VVh} The three 1-loop diagrams with an insertion of the effective operator $O_6$ that contribute to the $VVh$ vertex at ${\cal O}(\lambda)$. Here $\phi$ denotes the relevant would-be Goldstone field that needs to be included if the calculation is performed in a $R_\xi$ gauge. }
\end{center}
\end{figure}

The full ${\cal O}(\lambda)$ corrections to the renormalised $V V h$ vertex  thus arise  from the  1-loop diagrams  shown in Figure~\ref{fig:VVh} and a  tree-level counterterm graph involving a Higgs wave function renormalisation. We determine the relevant contributions using {\tt FeynArts}~\cite{Hahn:2000kx} and {\tt FormCalc}~\cite{Hahn:1998yk}.  Including the SM tree-level contribution, our final result for the renormalised $VVh$ vertex reads \begin{equation} \label{eq:threepoint}
\Gamma^{\mu \nu}_V (q_1, q_2) = 2 \left ( \sqrt{2} G_F \right )^{1/2} m_V^2 \,  \Big [ \eta^{\mu \nu} \, \left ( 1+ {\cal F} _1 (q_1^2, q_2^2) \right ) +  q_1^\nu q_2^\mu \; {\cal F}_2 (q_1^2, q_2^2)  \Big ] \,,
\end{equation}
where $G_F = 1/(\sqrt{2} \hspace{0.25mm} v^2)$ is the Fermi constant, $\eta^{\mu \nu}$ is the metric tensor, while  $m_V$ and $q_{i}^\mu$  with $i = 1,2$ denote the mass and the 4-momenta of the  external gauge bosons. The indices and momenta are assigned to the vertex as $V^\mu (q_1) + V^\nu (q_2) \to h(q_1 + q_2)$ with $(q_1 + q_2)^2 = m_h^2$,~i.e.~an on-shell Higgs boson. Notice that $\Gamma^{\mu \nu}_V (q_1, q_2)$ contains only Lorentz structures that gives rise to a non-vanishing contribution when the vertex is contracted with massless fermion lines, which is equivalent to including only transversal gauge-boson polarisations $\varepsilon_{i}^\mu (q_{i})$ in an on-shell calculation by requiring $\varepsilon_i (q_i) \cdot q_i  = 0$.

The form factors entering (\ref{eq:threepoint}) can be expressed in terms of the following 1-loop Passarino-Veltman (PV) scalar integrals 
\begin{equation} \label{eq:integrals1}
\begin{split}
& \hspace{2cm} B_0 \! \left (p_1^2, m_0^2, m_1^2 \right )  = \frac{\mu^{4-d}}{i \pi^{d/2} \hspace{0.25mm} r_\Gamma} \int \! \frac{d^d l}{\prod_{i=0,1} P (l+p_i,m_i)} \,, \\[2mm]
& \hspace{2.5cm}  B_0^\prime \! \left (p_1^2, m_0^2, m_1^2 \right )  = \left . \frac{\partial B_0 \! \left (k^2, m_0^2, m_1^2 \right ) }{\partial k^2} \right |_{k^2 \hspace{0.25mm} = \hspace{0.25mm} p_1^2} \,, \\[2mm]
&C_0 \! \left (p_1^2, (p_1-p_2)^2, p_2^2, m_0^2, m_1^2,m_2^2 \right )  = \frac{\mu^{4-d}}{i \pi^{d/2} \hspace{0.25mm} r_\Gamma} \int \! \frac{d^d l}{\prod_{i=0,1,2} P (l+p_i,m_i)} \,, 
\end{split}
\end{equation}
and the tensor coefficients of the two tensor integrals
\begin{equation} \label{eq:integrals2}
\begin{split}
C^\mu \! \left (p_1^2, (p_1-p_2)^2, p_2^2, m_0^2, m_1^2,m_2^2 \right ) & = \frac{\mu^{4-d}}{i \pi^{d/2} \hspace{0.25mm} r_\Gamma} \int \! \frac{d^d l \; l^\mu}{\prod_{i=0,1,2} P (l+p_i,m_i)} \,, \\[2mm]
C^{\mu \nu} \! \left (p_1^2, (p_1-p_2)^2, p_2^2, m_0^2, m_1^2,m_2^2 \right ) & = \frac{\mu^{4-d}}{i \pi^{d/2} \hspace{0.25mm} r_\Gamma} \int \! \frac{d^d l \; l^\mu \hspace{0.25mm} l^\nu}{\prod_{i=0,1,2} P (l+p_i,m_i)} \,. 
\end{split}
\end{equation}
Here $\mu$ is the renormalisation scale that keeps track of the correct dimension of the integrals in $d = 4 - 2 \hspace{0.25mm} \epsilon$ space-time dimensions, $r_\Gamma = \Gamma^2 (1-\epsilon) \hspace{0.25mm} \Gamma ( 1 + \epsilon)/\Gamma(1-2\epsilon)$ with $\Gamma(z)$ denoting the Euler gamma function, $P(k,m)=k^2 - m^2$ and $p_0 = 0$. The definitions (\ref{eq:integrals1})  and (\ref{eq:integrals2})  resemble those  of the  {\tt LoopTools} package~\cite{Hahn:1998yk}. 

The integrals with a tensor structure (\ref{eq:integrals2}) can be reduced to linear combinations of Lorentz-contravariant tensors constructed from the metric tensor $\eta^{\mu \nu}$ and a linearly independent set of the 4-momenta $p_i^\mu$. We define the tensor coefficients of the triangle integrals in the following way 
\begin{equation} \label{eq:tensors}
\begin{split}
C^\mu = \sum_{i=1,2} p_i^\mu \hspace{0.25mm} C_i \,, \qquad 
C^{\mu \nu} = \eta^{\mu \nu}  \hspace{0.25mm}   C_{00} + \sum_{i,j=1,2} p_i^\mu  p_j^\nu \hspace{0.25mm} C_{ij} \,.
\end{split}
\end{equation}
Notice that of all scalar and tensor-coefficient functions appearing in our 1-loop calculations only~$B_0$ and $C_{00}$ are ultraviolet (UV) divergent. These divergent contributions appear in our final results always in the UV-finite combination $B_0  - 4C_{00}$.  

With the definitions (\ref{eq:integrals1}), (\ref{eq:integrals2})  and (\ref{eq:tensors}) at hand, the full analytic expressions of the form factors can be written as 
\begin{equation} \label{eq:f1f2}
\begin{split}
{\cal F}_1 (q_1^2,q_2^2)& = \frac{\lambda \hspace{0.25mm} \bar c_6}{(4 \pi)^2} \left ( -3 B_0  - 12 \left ( m_V^2 \hspace{0.25mm} C_0 - C_{00}  \right ) - \frac{9 \hspace{0.25mm} m_h^2}{2} \left ( \bar c_6 + 2 \right ) B_0^\prime  \right ) \,, \\[2mm]
{\cal F}_2 (q_1^2,q_2^2)& =  \frac{\lambda \hspace{0.25mm} \bar c_6}{(4 \pi)^2}  \, 12 \, \big (  C_1 + C_{11} + C_{12} \big ) \,.
\end{split}
\end{equation}
Here the arguments of the PV integrals are 
\begin{equation} \label{eq:VVhnotation}
B_0 = B_0 \big (m_h^2,m_h^2,m_h^2 \big ) \,, \qquad 
C_0 = C_0 \big (m_h^2, q_1^2, q_2^2, m_h^2, m_h^2, m_V^2 \big ) \,,
\end{equation}
and analog definitions hold for  the derivative $B_0^\prime$ of the scalar bubble  integral and the tensor coefficients $C_1$, $C_{11}$ and $C_{12}$ of the triangle integral. Notice  that in contrast to \cite{Degrassi:2016wml} an all-order resummation of 1-loop wave function effects is not performed in~(\ref{eq:f1f2}). Since already the ${\cal O} (\lambda^2)$ wave function corrections in the SMEFT will be incomplete due to missing 2-loop Higgs-boson selfenergy  diagrams,  it is questionable if such a resummation  improves the precision of the calculation and we therefore do not  include it  our work. 

\section{Corrections to the Higgs partial decay widths}
\label{sec:widths}

\begin{figure}[!t]
\begin{center}
\includegraphics[width=0.975\textwidth]{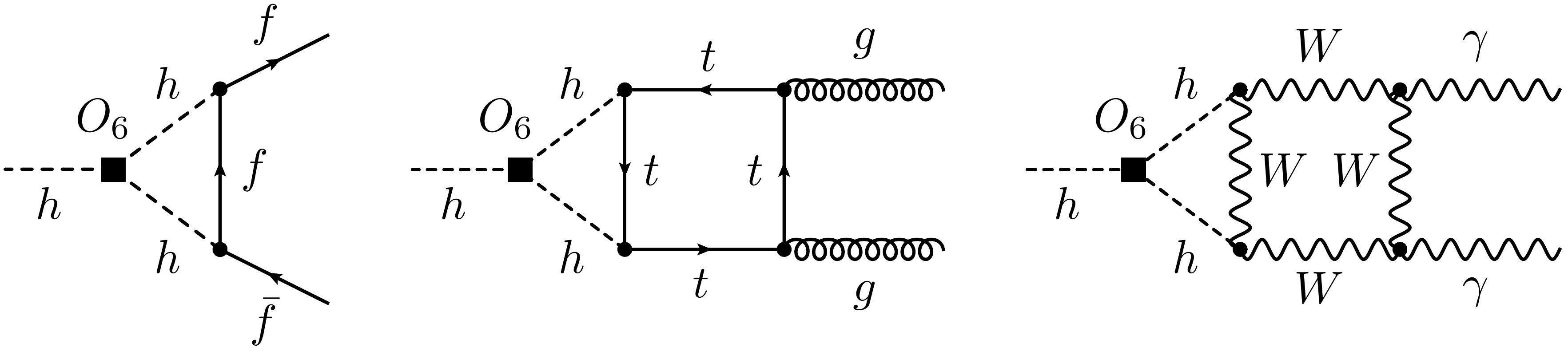}  
\vspace{2mm}
\caption{\label{fig:decays} Feynman diagrams with an insertion of the effective operator $O_6$ that lead to Higgs-boson decays into fermion (left), gluon (middle) and photon (right) pairs. }
\end{center}
\end{figure}

To determine the signal strengths in $Vh$ and VBF~Higgs production, one also has to take into account that the Higgs branching ratios are modified at the  loop level by the presence of the dimension-6 operator $O_6$.  Examples of diagrams that alter  the partial widths of the Higgs  to fermions, gluons and photons are displayed in Figure~\ref{fig:decays}. Below we will present results for the ${\cal O} (\lambda)$ corrections to the partial widths of all relevant Higgs decay modes. Terms of ${\cal O} (\lambda^2)$ that arise from squared matrix elements with an $O_6$ insertion are instead dropped for consistency since such contributions receive additional but unknown corrections from the interference of tree-level SM and loop-level SMEFT amplitudes. 

In the case of the decays of the Higgs to light fermion pairs $f = q, \ell$, we write
\beq \label{eq:DGff}
\Delta \Gamma (h \to f \bar f)  = \frac{N_c^f  G_F  \hspace{0.25mm} m_h \hspace{0.25mm} m_f^2}{4 \sqrt{2} \pi} \left ( 1 - \frac{4 \hspace{0.1mm} m_f^2}{m_h^2} \right )^{3/2}\, \Delta_f \,,
\eeq
where $N_c^q = 3$, $N_c^\ell = 1$ and all quark masses $m_q$ are understood as $\overline{\rm MS}$ masses renormalised at the scale $m_h$, while $m_\ell$ denotes the pole mass of the corresponding lepton. The ${\cal O} (\lambda)$ correction to the partial decay width $\Gamma (h \to f \bar f)$ stem from the graph displayed on the left-hand side in~Figure~\ref{fig:decays}. We obtain
\beq
\Delta_f  = \frac{\lambda \hspace{0.25mm}  \bar c_6}{(4 \pi)^2} \, {\rm Re} \, \Big (  - 12 \hspace{0.25mm} m_f^2 \left ( C_{0} - C_{1} - C_{2} \right )  -9 \hspace{0.25mm} m_h^2 \left ( \bar c_6 + 2 \right ) B_0^\prime \Big )   \,, \
\eeq
with 
\beq
C_0 = C_0 \big (m_f^2, m_h^2, m_f^2, m_f^2, m_h^2, m_h^2 \big ) \,,
\eeq 
and analogue definitions for the tensor coefficients $C_1$ and $C_2$. Notice that the flavour-dependent contributions are suppressed by light-fermion masses compared to the flavour-independent contribution proportional to $B_0^\prime$ that arises from the wave function renormalisation of the Higgs boson. The corrections $\Delta_f$ are hence to very good approximation universal. The result (\ref{eq:DGff})  agrees  numerically with \cite{Degrassi:2016wml}.

The shifts in the partial width for a Higgs boson decaying into a pair of EW gauge bosons can be cast into the form~\cite{Grau:1990uu}
\bea \label{eq:4body}
\Delta \Gamma (h \to V V) = \frac{1}{\pi^2} \int_0^{m_h^2} \! \frac{dq_1^2 \hspace{0.5mm} m_V \Gamma_V}{(q_1^2 - m_V^2)^2 + m_V^2 \Gamma_V^2}
\int_0^{(m_h -q_1)^2} \! \frac{dq_2^2 \hspace{0.5mm} m_V \Gamma_V}{(q_2^2 - m_V^2)^2 + m_V^2 \Gamma_V^2} \; I_{V} \,, \hspace{5mm}
\eea
and include  the contributions from both the production of one real and one virtual EW gauge boson $h \to V V^\ast$ or two virtual states $h \to V^\ast V^\ast$. In (\ref{eq:4body})  the total decay width of the relevant gauge boson is denoted by $\Gamma_V$ and the integrand can be written as 
\beq
I_{V} = \frac{G_F \hspace{0.25mm} m_h^3}{8 \sqrt{2} \pi} \, N_V \, \sqrt{\alpha (q_1^2, q_2^2, m_h^2)} \, \beta (q_1^2, q_2^2, m_h^2) \, \Delta_V \,, 
\eeq
with $N_W = 1$,  $N_Z = 1/2$ and 
\beq  \label{eq:alphabeta}
\alpha (x,y,z)  = \left ( 1 - \frac{x}{z} - \frac{y}{z} \right)^2 - \frac{4 \hspace{0.25mm} x y}{z^2} \,, \qquad 
\beta  (x,y,z)=  \alpha (x,y,z) + \frac{12 \hspace{0.25mm}  x y}{z^2} \,.
\eeq
The ${\cal O} (\lambda)$ correction to the partial decay width  $\Gamma (h \to V V)$ arises from the diagrams shown in Figure~\ref{fig:VVh}. We find
\beq
\begin{split}
\Delta_V & = \frac{\lambda \hspace{0.25mm}  \bar c_6}{(4 \pi)^2} \, {\rm Re} \Bigg [  -6 B_0   - 24 \left ( m_V^2 \hspace{0.25mm} C_0 - C_{00}\right ) \\[1mm]
& \phantom{xx}  - \frac{12  \hspace{0.25mm} \alpha (q_1^2, q_2^2, m_h^2) \left ( q_1^2 + q_2^2 - m_h^2 \right )   }{\beta (q_1^2, q_2^2, m_h^2)} \, \big (  C_1 + C_{11} + C_{12} \big ) - 9 \hspace{0.25mm} m_h^2 \left ( \bar c_6 + 2 \right ) B_0^\prime \Bigg ] \,.
\end{split}
\eeq
Here the arguments of the PV loop integrals are defined as in (\ref{eq:VVhnotation}).  We have verified that the expression (\ref{eq:4body})  agrees  numerically with the results presented in \cite{Degrassi:2016wml}.

The changes in partial decay widths of the Higgs boson to gluon and photon pairs can be written in the following way 
\beq 
\begin{split}
\Delta \Gamma ( h \to gg ) & = \frac{G_F \hspace{0.25mm} \alpha_s^2  \hspace{0.25mm} m_h^3}{36 \sqrt{2} \pi^3} \, \bigg | \sum_q A_q \hspace{0.5mm} \bigg |^2 \, \Delta_g  \,, \\[2mm]
\Delta \Gamma ( h \to \gamma \gamma ) & = \frac{G_F \hspace{0.25mm} \alpha^2  \hspace{0.25mm} m_h^3}{128 \sqrt{2} \pi^3} \, \bigg | \sum_f \frac{4 \hspace{0.25mm} N_c^f   Q_f^2}{3}  \, A_f - A_W  \hspace{0.25mm} \bigg |^2 \, \Delta_\gamma  \,, 
\end{split}
\eeq
where $\alpha_s = \alpha_s (m_h)$, $\alpha = 1/137.04$, while $Q_u=2/3$, $Q_d=-1/3$ and $Q_\ell =-1$ denote the electric charges of the fermions.
The leading-order (LO) form factors that encode the 1-loop corrections due to SM fermion and $W$-boson loops read 
\beq \label{eq:Af}
\begin{split}
A_f & = \frac{3 \hspace{0.25mm} \tau_f}{2} \left [ 1 + ( 1- \tau_f) \arctan^2 \frac{1}{\sqrt{\tau_f - 1}} \right ] \,, \\[3mm]
A_W & = 2+ 3\tau_W + 3 \tau_W (2-\tau_W) \arctan^2 \frac{1}{\sqrt{\tau_W-1}}  \,,
\end{split}
\eeq
with $\tau_X = 4 m_X^2/m_h^2$ for $X = f, W$.  The ${\cal O} (\lambda)$ correction to the partial decay width of the Higgs to gluons and photons originate from 2-loop diagrams with an insertion of $O_6$. Two example graphs are shown in the middle and on the right of Figure~\ref{fig:decays}.  The results presented  in \cite{Gorbahn:2016uoy,inprep} lead to 
\beq 
\begin{split}
\Delta_g & = \frac{\lambda \hspace{0.25mm}  \bar c_6}{(4 \pi)^2}  \, \big (  8.42 - 9 \hspace{0.25mm} m_h^2 \left ( \bar c_6 + 2 \right ) B_0^\prime \big  )    \,, \\[2mm]
\Delta_\gamma & = \frac{\lambda \hspace{0.25mm}  \bar c_6}{(4 \pi)^2}\, \big (  -3.70 - 9 \hspace{0.25mm} m_h^2 \left ( \bar c_6 + 2 \right ) B_0^\prime \big  )     \,.
\end{split}
\eeq 
Notice that there is no need to take the real part here because the $B_0^\prime$ integral corresponding to a Higgs loop is real for on-shell kinematics.  The expression for $\Delta_g$ agrees  with the results obtained in \cite{Degrassi:2016wml}.

\section{Description of the $\bm{Vh}$ calculation}
\label{sec:Vh}

In order to explain how we obtain our predictions for the associated production of the Higgs boson  with massive gauge bosons it is useful to first consider the ${\cal O} (\lambda)$ corrections to~$\sigma_{Vh} = \sigma (q \bar q \to V h)$ working to zeroth order in the strong coupling constant. At this order in~QCD the ${\cal O} (\lambda)$ shift in the integrated partonic cross section can be written as  
\begin{equation}
\Delta \sigma_{Vh} = \frac{G_F^2 \hspace{0.1mm}  m_V^4}{72 \pi \hspace{0.25mm}} \, \bar N_V  \, \sqrt{\alpha (m_V^2, m_h^2, s)} \,\, \frac{\alpha (m_V^2, m_h^2, s) \, s+ 12 \hspace{0.25mm} m_V^2}{\left (s -m_V^2 \right )^2} \; \delta_V  \,,
\end{equation}
with $\bar N_W =1$ and $\bar N_Z =  \left ( 1 - 8 \hspace{0.25mm} T_3^q \hspace{0.25mm} Q_q s_w^2 + 8 \hspace{0.25mm} Q_q^2 s_w^4 \right)/2$, where  $T_3^q$ ($Q_q)$ denotes the third component of the weak isospin (electric charge) of the relevant quark. The function $\delta_V$ encodes the contributions from the three 1-loop diagrams in Figure~\ref{fig:VVh} when one of the gauge bosons is contracted with a quark line and the other one is put on its mass shell. Explicitly we find 
\begin{equation} \label{eq:Vhcorrection}
\begin{split}
\delta_V  & = \frac{\lambda \hspace{0.25mm}  \bar c_6}{(4 \pi)^2} \, {\rm Re} \Bigg [  -6 B_0   - 24 \left ( m_V^2 \hspace{0.25mm} C_0 - C_{00}\right ) \\[1mm]
& \phantom{xx}  - \frac{12  \hspace{0.25mm} \alpha (m_V^2, m_h^2, s) \, s \left (m_V^2 -m_h^2 + s \right )   }{ \alpha (m_V^2, m_h^2, s)  \, s  +12 \hspace{0.25mm} m_V^2} \, \big (  C_1 + C_{11} + C_{12} \big ) - 9 \hspace{0.25mm} m_h^2 \left ( \bar c_6 + 2 \right ) B_0^\prime \Bigg ] \,,
\end{split}
\end{equation}
where the function $\alpha (x,y,z)$   has  been defined in (\ref{eq:alphabeta}).  The arguments of the scalar triangle integral are 
\beq
C_0 = C_0 \big (m_h^2, s, m_V^2, m_h^2, m_h^2, m_V^2 \big ) \,,
\eeq
and all other  tensor coefficients carry the same functional dependence. The $B_0$ integral is defined in (\ref{eq:VVhnotation}). Our result (\ref{eq:Vhcorrection}) for~$\delta_V$ can be shown to agree with the analytic expression given in the publication \cite{McCullough:2013rea} for the case of~$e^+ e^- \to Zh$. 

\begin{figure}[!t]
\begin{center}
\includegraphics[width=0.95\textwidth]{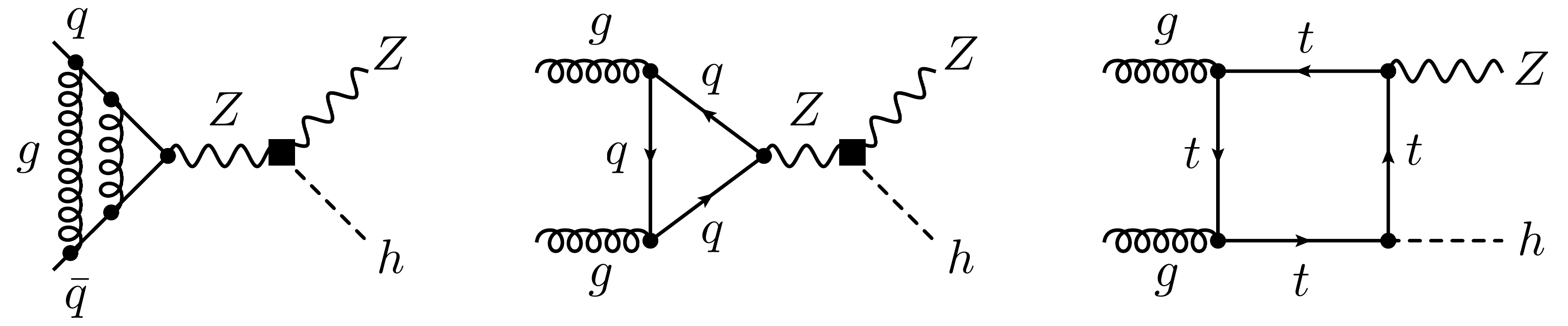}  
\vspace{2mm}
\caption{\label{fig:Vh}  Examples of diagrams that contribute  to $p p \to Zh$ at ${\cal O} (\alpha_s^2)$.  As indicated by the black square the left and middle diagram receive a correction of ${\cal O} (\lambda)$ from $\delta_V$,  while the graph on the right-hand side does not involve a modified $ZZh$ vertex. See text for further explanations.}
\end{center}
\end{figure}

At NNLO the production cross section for $pp \to Vh$  receives corrections from two types of topologies. The first  kind of graphs involves an exchange of a single off-shell vector boson in the $s$-channel, while the second sort of corrections arise from the coupling of the Higgs boson to a closed loop of top quarks. For on-shell bosons the former type of~${\cal O} (\alpha_s^2)$ corrections have been obtained in \cite{Brein:2003wg}, while fully differential NNLO calculations of these Drell-Yan~(DY) parts have been presented  in  \cite{Ferrera:2011bk,Ferrera:2013yga} and \cite{Ferrera:2014lca} for the $Wh$ and~$Zh$ final state, respectively. Subsets of the diagrams where the Higgs is radiated off a top loop have been considered in \cite{Brein:2011vx,Ferrera:2014lca} and a calculation of all such graphs can be found in~\cite{Campbell:2016jau}. The latter results have been implemented into version 8 of {\tt MCFM}~\cite{Boughezal:2016wmq}.

The existing fully differential {\tt MCFM} implementation of $pp \to Vh$ at NNLO serves as a starting point of our own  computation.    We have identified the  routines in~{\tt MCFM} that correspond to the two different kinds of ${\cal O}(\alpha_s^2)$ corrections. For the case of $pp \to Zh$ representatives of the two types of contributions are displayed in~Figure~\ref{fig:Vh}. Notice that in all diagrams where the Higgs is not radiated from a top loop the $\delta_V$ correction factorises and thus we are able to include the complete~${\cal O} (\lambda)$ term (\ref{eq:Vhcorrection}) on top of the~NNLO corrections.  In the case of the contributions with top loops however not all ${\cal O} (\alpha_s^2)$ corrections factorise. Non-factorisable contributions which involve a top box and a top-Higgs triangle as well as double-box contributions  are in fact not known and thus cannot be included. Effects due to Higgs wave function renormalisation, on the other hand, factorise and we take them into account in our computations. As a result, our numerical predictions for the differential  $pp \to Vh$  cross sections are NNLO accurate only for what concerns the ${\cal O} (\lambda)$ terms associated to $\delta_V$, while we are missing ${\cal O} (\alpha_s^2)$ contributions proportional to $\lambda \bar c_6$ that stem from top loops.

\section{Description of the VBF~Higgs calculation}
\label{sec:VBF}

To obtain predictions for VBF~Higgs production we employ the structure-function approach~\cite{Han:199hhr}. In this formalism the  VBF~Higgs process can be described to high  accuracy as a double deep-inelastic scattering process (DIS), where two virtual EW gauge bosons emitted from the hadronic initial states fuse into a Higgs boson.  Neglecting small QCD-interference effects between the two inclusive final states, the differential VBF~Higgs cross section is in our case given by a product of two 3-point vertices $\Gamma^{\mu \nu}_V (Q_1,Q_2)$ and two  DIS hadronic tensors $W^{\mu \nu}_V (x_i, Q_i^2)$:
\begin{equation} \label{eq:dsigma}
\begin{split}
d\sigma_{\rm VBF} & = \frac{G_F^2 \hspace{0.1mm} m_V^4}{s} \hspace{0.5mm} \Delta_V^2 (Q_1^2)  \hspace{0.25mm}   \Delta_V^2 (Q_2^2)  \\[2mm] & \phantom{xx} \times W_{\mu \nu}^V (x_1, Q_1^2) \hspace{0.5mm}  \Gamma^{\mu \rho}_V  (Q_1, Q_2) \hspace{0.25mm}  \big ( \Gamma^{\nu \sigma} _V (Q_1, Q_2) \big  )^\ast \hspace{0.5mm} W_{\rho \sigma}^V (x_2, Q_2^2) \, d\Omega \,.
\end{split}
\end{equation}
Here $\Delta_V (Q_{i}^2) = 1/(Q_{i}^2 + m_V^2)$, $Q_{i}^2 = -q_{i}^2$ and $x_{i} = Q^2_{i}/(2P_{i} \cdot q_{i})$ are the usual~DIS variables  with $P_i^\mu$ the 4-momentum of proton $i =1,2$ and $d\Omega$  denotes the 3-particle VBF phase space. The hadronic tensor can be expressed as 
\begin{equation} \label{eq:WmunuV} 
\begin{split} 
 W^{\mu \nu}_V (x_i, Q_i^2) &  = \left ( -\eta^{\mu \nu} - \frac{q_i^\mu q_i^\nu}{Q_i^2} \right ) F_1^V (x_i, Q_i^2) + \frac{\hat P_i^\mu \hat P_i^\nu}{P_{ii}} \, F_2^V (x_i, Q_i^2) \\[2mm]  & \phantom{xx} + i \epsilon^{\mu \nu \rho \sigma} \, \frac{P_{i \hspace{0.25mm} \rho} \hspace{0.25mm} q_{i \hspace{0.25mm} \sigma}}{2 P_{ii}}  \, F_3^V (x_i, Q_i^2) \,,
\end{split}
\end{equation}
where $\epsilon^{\mu \nu \rho \sigma}$ is the fully anti-symmetric Levi-Civita tensor and  we have introduced 
\begin{equation}
\qquad P_i \cdot q_j = P_{ij} \,, \qquad \hat P_i^\mu = P_i^\mu + \frac{P_{ii}}{Q_i^2} \, q_i^\mu \,.
\end{equation}
The standard DIS structure functions are denoted by $F_m^V (x_i,  Q_i^2)$ with $m = 1,2,3$. 

Using the decomposition (\ref{eq:WmunuV}) the squared hadronic tensor in (\ref{eq:dsigma}) can be written in terms of the DIS structure functions as 
\begin{equation}
\begin{split}
W_{\mu \nu}^V (x_1, Q_1^2) \hspace{0.5mm}  \Gamma^{\mu \rho}_V  (Q_1,Q_2) \hspace{0.25mm}  \big ( \Gamma^{\nu \sigma}_V  (Q_1,Q_2) \big  )^\ast \hspace{0.5mm} W_{\rho \sigma}^V (x_2, Q_2^2) & =  \\[2mm] & \hspace{-3cm} 4 \sqrt{2} \hspace{0.25mm} G_F \hspace{0.25mm}  m_V^4 \sum_{m,n}^3 w_{mn} \hspace{0.25mm} F_m^V(x_1, Q_1^2)  \hspace{0.25mm} F_n^V (x_2, Q_2^2)  \,.  
\end{split}
\end{equation}
Defining the short-hand notations\footnote{For VBF kinematics the form factor ${\cal F}_{1,2} (Q_1^2, Q_2^2)$ are real and in consequence there is no need to take the real part in the last two definitions in (\ref{eq:def}).}
\begin{equation} \label{eq:def}
q_1 \cdot q_2 = q_{12} \,, \quad P_1 \cdot P_2 = p_{12}  \,, \quad  {\cal C}_1 = 1 + 2 \hspace{0.25mm}  {\cal F}_1 (Q_1^2, Q_2^2) \,, \quad {\cal C}_2 = 2 \hspace{0.25mm}  {\cal F}_2 (Q_1^2, Q_2^2) \,, 
\end{equation}
the non-vanishing coefficients $w_{mn}$ included in our analysis read  
\begin{eqnarray} \label{eq:Cmn}
\begin{split}
w_{11} & = \left ( 2 \hspace{0.25mm} {\cal C}_1  -   q_{12}  \hspace{0.5mm} {\cal C}_2 \right ) + \left (  {\cal C}_1 +  q_{12}  \hspace{0.5mm} {\cal C}_2 \right ) \hspace{0.25mm} \frac{q_{12}^2}{Q_1^2 \hspace{0.25mm} Q_2^2}  \,, \\[2mm]
w_{12} & = -{\cal C}_1 \hspace{0.25mm}  \frac{P_{22}}{Q_2^2}  - \left ( {\cal C}_1  + q_{12}  \hspace{0.5mm}  {\cal C}_2  \right ) \left ( \frac{P_{21}^2}{P_{22} \hspace{0.25mm} Q_1^2} + \frac{2 P_{21} \hspace{0.5mm} q_{12}}{Q_1^2 \hspace{0.25mm} Q_2^2} + \frac{P_{22} \hspace{0.5mm}  q_{12}^2}{Q_1^2 \hspace{0.25mm} Q_2^4} \right )    \,, \\[2mm]
w_{21} & =  -{\cal C}_1 \hspace{0.25mm}  \frac{P_{11}}{Q_1^2}  - \left ( {\cal C}_1  + q_{12}  \hspace{0.5mm} {\cal C}_2  \right ) \left ( \frac{P_{12}^2}{P_{11} \hspace{0.25mm} Q_2^2} + \frac{2 P_{12} \hspace{0.5mm} q_{12}}{Q_1^2 \hspace{0.25mm} Q_2^2} + \frac{P_{11} \hspace{0.5mm}  q_{12}^2}{Q_1^4 \hspace{0.25mm} Q_2^2} \right )   \,, \\[2mm]
w_{22} & = \frac{1}{P_{11} P_{22} \hspace{0.5mm} Q_1^4 \hspace{0.25mm} Q_2^4 } \, \Bigg [ \, {\cal C}_1\, \Big ( p_{12} \hspace{0.5mm} Q_1^2 Q_2^2 + P_{11} P_{21} \hspace{0.25mm} Q_2^2 + P_{12} P_{22} \hspace{0.25mm} Q_1^2  + P_{11} P_{22} \hspace{0.5mm} q_{12} \Big )^2  \\[1mm]
& \phantom{xx}  + {\cal C}_2 \left (P_{12} \hspace{0.25mm} Q_1^2 + P_{11} \hspace{0.5mm}  q_{12} \right ) \left (P_{21} \hspace{0.25mm}  Q_2^2 +P_{22} \hspace{0.5mm}  q_{12} \right ) \\[1mm] 
& \phantom{xxxi} \times \Big  (p_{12} \hspace{0.5mm} Q_1^2 Q_2^2 + P_{11} P_{21} \hspace{0.25mm}  Q_2^2+ P_{12} P_{22} \hspace{0.25mm} Q_1^2 +P_{11} P_{22} \hspace{0.5mm}  q_{12} \Big  )  \Bigg ] \,, \\[2mm]
 w_{33} & = \frac{1}{4 P_{11} P_{22}} \, \bigg [ \, 2 \hspace{0.25mm} {\cal C}_1 \left ( p_{12} \hspace{0.5mm} q_{12} - P_{12}  \hspace{0.25mm} P_{21} \right ) \\[1mm] &  \phantom{xx} - {\cal C}_2 \, \Big \{ P_{12} P_{22} \hspace{0.25mm} Q_1^2 + \left (p_{12} \hspace{0.5mm} Q_1^2 + P_{11} P_{21} \right ) Q_2^2 + \left (P_{11} P_{22} + P_{12} P_{21} \right ) q_{12} - p_{12} \hspace{0.5mm} q_{12}^2 \Big \} \, \bigg ] \,.
\end{split}
\end{eqnarray}
Notice that in the above expressions for the coefficients $w_{mn}$ we have neglected terms quadratic in the form factors ${\cal F}_{1,2} (Q_1^2, Q_2^2)$. Such contributions are suppressed relative to the linear terms in (\ref{eq:Cmn}) by a factor of $\lambda \bar c_6/(4 \pi)^2$ and thus formally of  2-loop order in the SMEFT. Since the 2-loop SMEFT contributions to the ${\cal O} (\lambda^2)$ corrections remain unknown including terms quadratic in (\ref{eq:f1f2}) would thus not improve the accuracy of the calculation. 

With all the non-vanishing coefficients $w_{mn}$ at hand it is now rather straightforward to calculate NNLO QCD corrections to the inclusive~\cite{Bolzoni:2010xr, Bolzoni:2011cu} and exclusive~\cite{Cacciari:2015jma} VBF~Higgs cross section.\footnote{Very recently  the next-to-next-to-next-to-leading order (N$^3$LO) QCD corrections to the inclusive VBF~Higgs cross section have been calculated in the structure-function approach~\cite{Dreyer:2016oyx}. We do not include  N$^3$LO effects  in our analysis since they amount to ${\cal O} (1 \permil)$ shifts, which is well within the NNLO scale uncertainties.} Our computations rely on the techniques and the Monte~Carlo (MC) codes developed in the latter work. In the inclusive part of the calculation, we employ the phase space from the $h+2 \, {\rm jets}$ VBF calculation implemented in~{\tt POWHEG}~\cite{Nason:2009ai}, while the matrix element is evaluated with structure functions based on parametrised versions~\cite{vanNeerven:1999ca,vanNeerven:2000uj} of the NNLO~DIS coefficient functions \cite{vanNeerven:1991nn,Zijlstra:1992qd,Zijlstra:1992kj} integrated with {\tt HOPPET}~\cite{Salam:2008qg}. The exclusive calculation relies  also on the NLO part of the {\tt POWHEG} $h+3 \, {\rm jets}$ VBF code~\cite{Jager:2014vna}, which implements the results of \cite{Figy:2007kv}. To take into account contributions from the second Lorentz structure in~(\ref{eq:threepoint}) the SM implementation \cite{Jager:2014vna} had to be extended. This extension required, in particular, new tree-level $h+4 \, {\rm jets}$ matrix elements, which were generated with {\tt MadGraph5\_aMC\@NLO}~\cite{Alwall:2014hca}. The numerical evaluation of 1-loop Feynman integrals is performed by {\tt QCDLoop}~\cite{Ellis:2007qk, Carrazza:2016gav} after reducing the tensor coefficients appearing in (\ref{eq:integrals2}) to basic PV scalar integrals. Further technical details on the implementation of the NNLO VBF~Higgs cross section computations are given in~\cite{Cacciari:2015jma}. 

\section{Numerical results}
\label{sec:numerics}

In this section we study the numerical impact of the ${\cal O} (\lambda)$ corrections that we have derived  earlier in Sections~\ref{sec:VVh} and~\ref{sec:widths}. We first present results for the modifications of the  Higgs production cross sections $\sigma_I$ in the vector boson mediated channels  $I = Wh, Zh, {\rm VBF}$. Then we study the corrections to the partial Higgs decay widths $\Gamma^F = \Gamma (h \to F)$ and branching ratios~${\rm Br}^F = {\rm Br} (h \to F)$. This discussion is followed by an analysis of the shape changes in the $Vh$ and VBF Higgs distributions due to  the ${\cal O} (\lambda)$ corrections. We finally derive the constraints on the Wilson coefficient~$\bar c_6$ that arise from LHC Run I and II data, and explore the prospects of the~HL-LHC in improving the current bounds. Both the limits from double-Higgs production as well as $Vh$ and VBF Higgs production are considered. 

\subsection{Modifications of the Higgs production cross sections}
\label{sec:sigmaI}

\begin{figure}[!t]
\begin{center}
\includegraphics[width=0.925\textwidth]{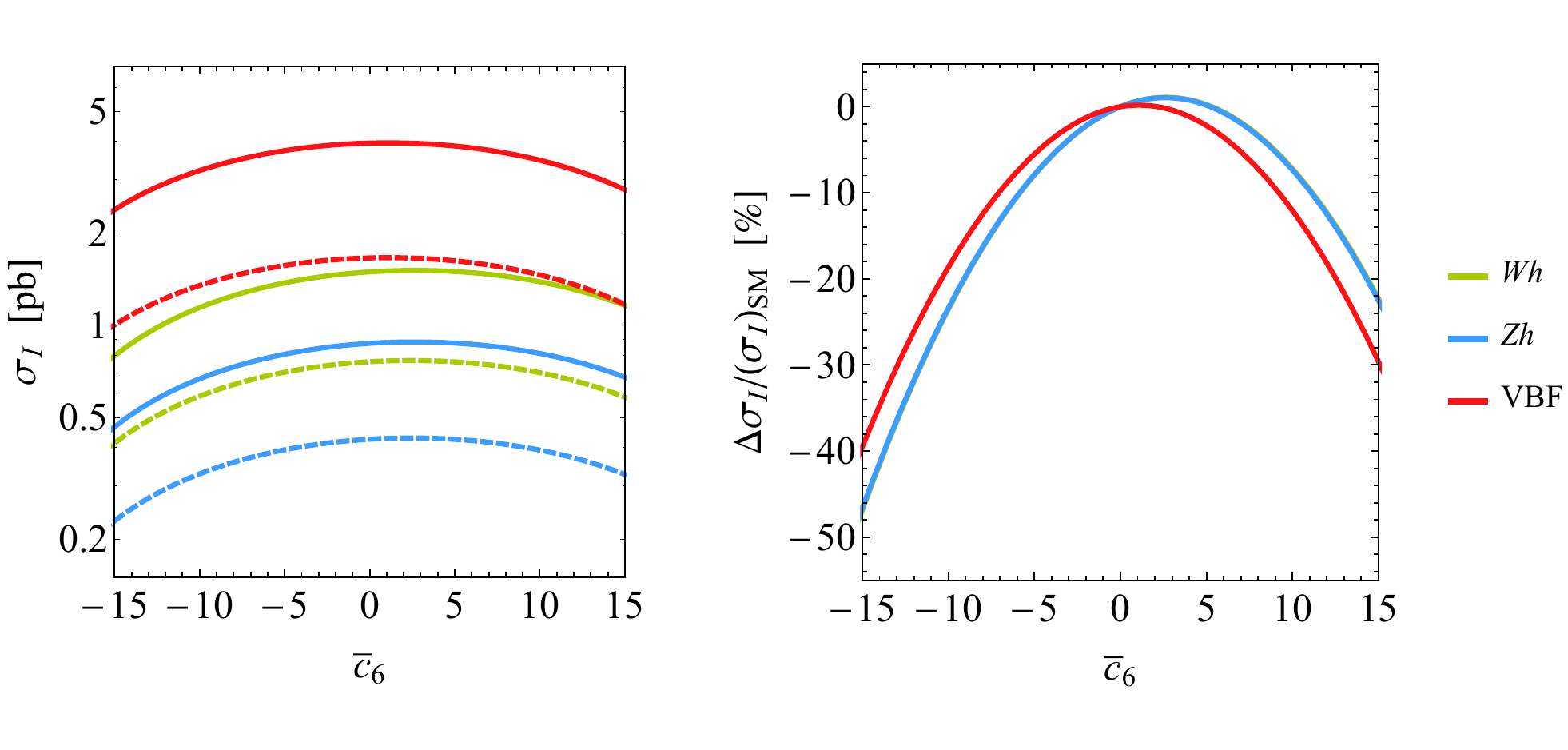}  
\vspace{-4mm}
\caption{\label{fig:sigmashifts}  Left: Predictions for the inclusive $Wh$, $Zh$ and VBF Higgs production sections $\sigma_I$ as a function of $\bar c_6$.  The dashed and solid curves correspond to $\sqrt{s} = 8 \, {\rm TeV}$ and $\sqrt{s} = 13 \, {\rm TeV}$, respectively. Right:~Relative modifications of $\sigma_I$ as a function of the Wilson coefficient of  $O_6$. Only results for $\sqrt{s} = 13 \, {\rm TeV}$ are shown.}
\end{center}
\end{figure}

We begin our discussion by considering  the modifications of the inclusive vector boson mediated  Higgs production sections~$\sigma_I$ that result from the presence of the ${\cal O} (\lambda)$ corrections. The corresponding predictions are shown in the two panels of Figure~\ref{fig:sigmashifts}  as a function of $\bar c_6$. In the left plot we display the total cross sections for $pp$ collisions at $\sqrt{s} = 8 \, {\rm TeV}$ (dashed curves) and $\sqrt{s} = 13 \, {\rm TeV}$~(solid curves). In the former case, we find  
\beq \label{eq:sigma8TeV}
\begin{split} 
\sigma_{Wh}^{8 \, {\rm TeV}} & = (\sigma_{Wh}^{8 \, {\rm TeV}} )_{\rm SM} \left ( 1 + 7.4  \cdot 10^{-3}  \hspace{1mm} \bar c_6 -1.5 \cdot 10^{-3}  \hspace{1mm}  \bar c_6^2 \right ) \,, \\[2mm]
\sigma_{Zh}^{8 \, {\rm TeV}}  & = (\sigma_{Zh}^{8 \, {\rm TeV}} )_{\rm SM} \left ( 1 + 7.5   \cdot 10^{-3}  \hspace{1mm} \bar c_6 -1.5 \cdot 10^{-3}  \hspace{1mm}  \bar c_6^2 \right ) \,, \\[2mm]
\sigma_{\rm VBF}^{8 \, {\rm TeV}}  & = (\sigma_{\rm VBF}^{8 \, {\rm TeV}} )_{\rm SM} \left ( 1 + 3.3 \cdot 10^{-3}  \hspace{1mm} \bar c_6 -1.5 \cdot 10^{-3}  \hspace{1mm} \bar c_6^2 \right ) \,,
\end{split}  
\eeq 
where the prediction for the $Wh$ cross section includes both the $pp \to W^+ h$ and the $pp \to W^- h$ channel. The SM predictions that enter the above formulas read $(\sigma_{Wh}^{8 \, {\rm TeV}} )_{\rm SM} =  ( 0.76 \pm 0.02) \, {\rm pb}$, $(\sigma_{Zh}^{8 \, {\rm TeV}} )_{\rm SM} =  ( 0.42 \pm 0.01) \, {\rm pb}$ and $(\sigma_{\rm VBF}^{8 \, {\rm TeV}} )_{\rm SM} =  ( 1.66 \pm 0.04 ) \, {\rm pb}$. In the latter case, we  instead obtain 
\beq \label{eq:sigma13TeV}
\begin{split} 
\sigma_{Wh}^{13 \, {\rm TeV}} & = (\sigma_{Wh}^{13 \, {\rm TeV}} )_{\rm SM} \left ( 1 + 8.2  \cdot 10^{-3}  \hspace{1mm} \bar c_6 -1.5 \cdot 10^{-3}  \hspace{1mm}  \bar c_6^2 \right ) \,, \\[2mm]
\sigma_{Zh}^{13 \, {\rm TeV}}  & = (\sigma_{Zh}^{13 \, {\rm TeV}} )_{\rm SM} \left ( 1 + 8.0   \cdot 10^{-3}  \hspace{1mm} \bar c_6 -1.5 \cdot 10^{-3}  \hspace{1mm}  \bar c_6^2 \right ) \,, \\[2mm]
\sigma_{\rm VBF}^{13 \, {\rm TeV}}  & = (\sigma_{\rm VBF}^{13 \, {\rm TeV}} )_{\rm SM} \left ( 1 + 3.3 \cdot 10^{-3}  \hspace{1mm} \bar c_6 -1.5 \cdot 10^{-3}  \hspace{1mm} \bar c_6^2 \right ) \,,
\end{split}  
\eeq 
and  the relevant~SM cross sections are $(\sigma_{Wh}^{13 \, {\rm TeV}} )_{\rm SM} =  ( 1.49 \pm 0.03) \, {\rm pb}$, $(\sigma_{Zh}^{13 \, {\rm TeV}} )_{\rm SM} = (  0.87 \pm 0.03) \, {\rm pb}$ and $(\sigma_{\rm VBF}^{13 \, {\rm TeV}} )_{\rm SM} = ( 3.94 \pm 0.08 ) \, {\rm pb}$.  Our results have been obtained with the implementations of the $Vh$ and VBF Higgs calculations described in Sections \ref{sec:Vh} and~\ref{sec:VBF}.  They correspond to {\tt PDF4LHC15\_nnlo\_mc} parton   distribution functions~(PDFs)~\cite{Butterworth:2015oua} and the quoted uncertainties include both scale, PDF and $\alpha_s$ errors. In the case of~$Vh$~(VBF Higgs) production our default scale choice is  $\mu_0 = m_V + m_h$~($\mu_0 = m_h$). The perturbative uncertainties  are estimated in both cases by identifying the renormalisation and factorisation scales $\mu_R$ and $\mu_F$ with $\mu_0$ and  varying $\mu_0$ by a factor of two around the default scale. 

The above formulas can be compared to the next-to-leading order~(NLO) results for the $Vh$ and VBF Higgs production cross sections presented in \cite{Degrassi:2016wml}. Concerning~$\sigma_{Wh}$ and $\sigma_{\rm VBF}$, we find that the inclusion of ${\cal O} (\alpha_s^2)$ corrections essentially does not change the  functional dependence  on $\bar c_6$ compared to NLO. In the case of~$\sigma_{Zh}$, NNLO effects have instead an impact since  they shift the term linear in~$\bar c_6$ by around $-20 \%$~($-10\%$) compared to the $8 \, {\rm TeV}$ ($13 \, {\rm TeV}$) NLO prediction. The observed shifts originate from the negative ${\cal O} (\alpha_s^2)$ contributions due to heavy-quark boxes of the type $gg \to Zh$. Given that the corresponding non-universal ${\cal O} (\lambda)$ corrections are not included in our calculation~(see the discussion in Section~\ref{sec:Vh}) it remains unclear whether the inclusion of NNLO effects improves the precision of our $pp \to Zh$ predictions. We  add that we have verified that at NLO our numerical results for $Vh$ and VBF Higgs production all agree  with the predictions given in~\cite{Degrassi:2016wml}. 

\begin{figure}[!t]
\begin{center}
\includegraphics[width=0.95\textwidth]{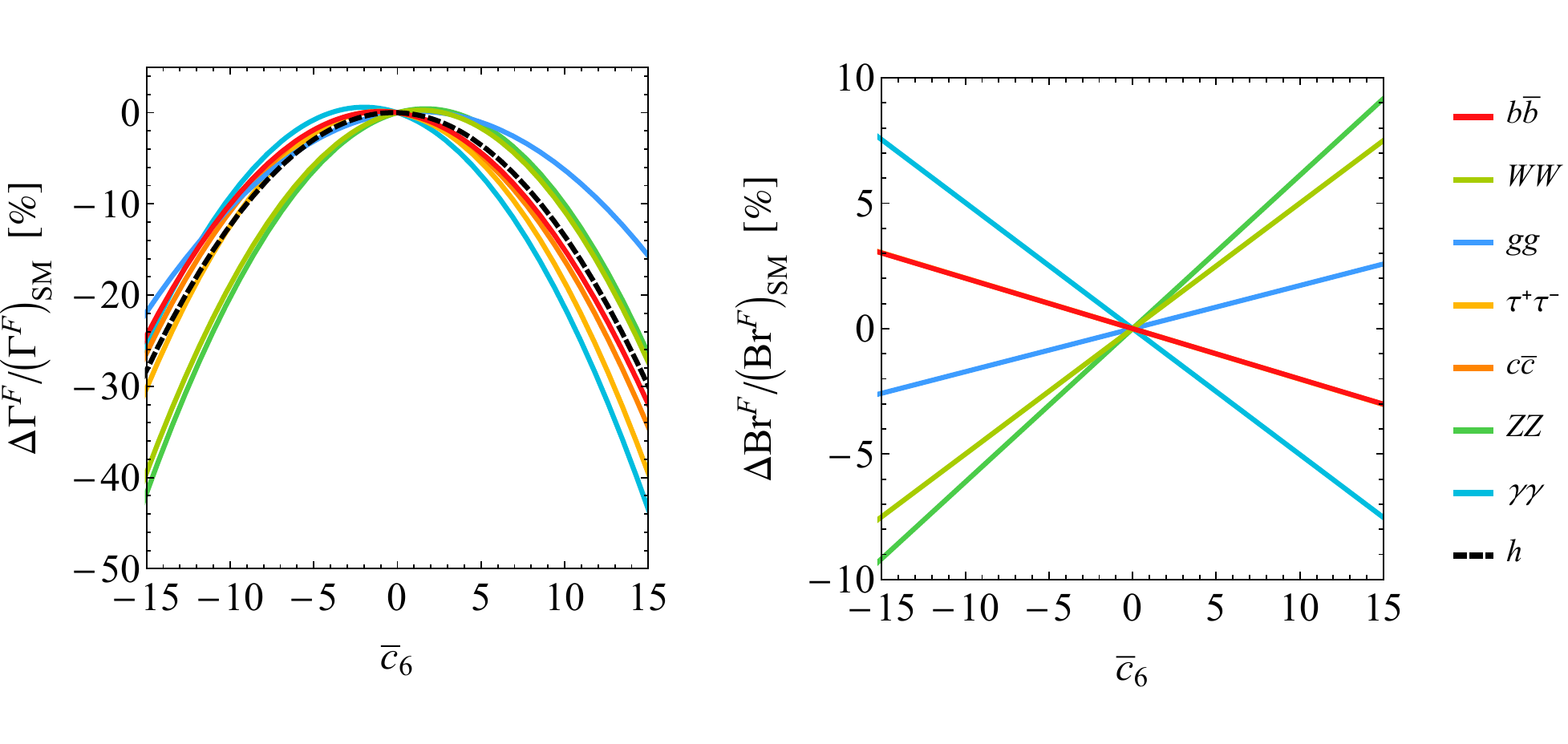}  
\vspace{-4mm}
\caption{\label{fig:gabrshifts} Shifts in the partial decay widths (left panel) and the branching ratios (right panel) of the Higgs boson as a function of the Wilson coefficient $\bar c_6$. The coloured curves indicate the individual decay channels, while the black dashed curve corresponds to the total Higgs decay width. }
\end{center}
\end{figure}

Looking at  the results (\ref{eq:sigma8TeV}) and (\ref{eq:sigma13TeV}) one  observes that the linear dependence on the Wilson coefficient $\bar c_6$ of the $Vh$ and~VBF Higgs cross sections is different. This feature is expected because the terms linear in $\bar c_6$  originate from both tree-level counterterm graphs involving a Higgs wave function renormalisation as well as the interference of tree-level with 1-loop amplitudes. While the Higgs wave function renormalisation constant depends only on $m_h$, the interference contributions have a non-trivial dependence on the external 4-momenta. As a result the ${\cal O} (\bar c_6)$ terms are process and kinematics dependent.  To better  illustrate the numerical impact of the~${\cal O} (\lambda)$ corrections, we plot $\Delta \sigma_I/(\Delta \sigma_I)_{\rm SM}$ as a function of $\bar c_6$ in the right panel of Figure~\ref{fig:sigmashifts} employing $\sqrt{s} = 13 \, {\rm TeV}$. We see that for $\bar c_6 \simeq -15$ the $Vh$ and VBF Higgs cross sections are shifted by about  $-50\%$  and $-40\%$, while for $\bar c_6 \simeq 15$ the corresponding shifts are around    $-25\%$ and $-30\%$. Given that  the functional dependencies of (\ref{eq:sigma8TeV}) and (\ref{eq:sigma13TeV}) are approximately the same, effects of similar size are obtained at $\sqrt{s} = 8 \, {\rm TeV}$. The corresponding predictions are not shown in the latter figure.

\subsection{Modifications of the Higgs decays}
\label{sec:BrF}

We now turn our attention to the partial decay widths  and branching ratios of the Higgs. In~Figure~\ref{fig:gabrshifts} we illustrate the numerical impact of the ${\cal O} (\lambda)$ corrections on these observables. As input parameters we have used $\alpha_s (m_h) = 0.1127$, $m_t = 173.2 \, {\rm GeV}$, $m_b (m_h) = 2.81 \, {\rm GeV}$, $m_c (m_h) = 0.65 \, {\rm GeV}$, $m_\tau = 1.777 \, {\rm GeV}$,  $m_W = 80.37 \, {\rm GeV}$, $m_Z = 91.15 \, {\rm GeV}$,   $\Gamma_W =   2.0886 \, {\rm GeV}$ and $\Gamma_Z =  2.4958 \, {\rm GeV}$. The quoted values for the bottom and charm quark $\overline{\rm MS}$ masses have been obtained by employing 2-loop running. The SM predictions for the total decay width of the Higgs and its branching ratios are taken from~\cite{YR4}. In the case of the partial decay widths (left panel), one observes that the relative corrections to $ \Gamma^F$  all have a very similar $\bar c_6$ dependence and are essentially always negative.  These features are related to the fact that for $|\bar c_6| \gtrsim 1$ the partial decay widths are dominated  by the universal corrections arising from the Higgs wave function renormalisation which is quadratic in $\bar c_6$ and carries a minus sign. Numerically, we find that the relative shifts in~$ \Gamma^F$ can reach up to around~$-40\%$~($-45\%$) for~$\bar c_{6} \simeq -15$~($\bar c_{6} \simeq 15$). The corrections to the total decay  width~$\Gamma_h$ are only about $-30\%$.  In the case of the shifts in the Higgs branching ratios (right panel), one observes instead that the modifications in all channels  do not exceed $\pm 10\%$ in the same~$\bar c_6$ range. The impact of ${\cal O} (\lambda)$ corrections   is thus generically smaller in the branching ratios than  in the partial decay widths, since in the former quantities the universal Higgs wave function corrections and thus the quadratic dependence on $\bar c_6$ cancels.

\subsection[Modifications of the $Vh$ and VBF Higgs distributions]{Modifications of the $\bm{Vh}$ and VBF Higgs distributions}
\label{sec:distributions}

Since the vertex corrections (\ref{eq:threepoint}) depend in a non-trivial way on the external 4-momenta, the ${\cal O} (\lambda)$ corrections not only change the overall size of the cross sections in $Vh$ and VBF Higgs production but also modify the shape of the corresponding kinematic distributions. In this subsection we present  results for the spectra that are most sensitive to modifications in the trilinear Higgs coupling. All results shown below correspond to $\sqrt{s} = 13 \, {\rm TeV}$, {\tt PDF4LHC15\_nnlo\_mc} PDFs and the default scale choices introduced in Section~\ref{sec:sigmaI}.  Off-shell effects in Higgs-boson production are taken into account by modelling the width of the Higgs with a  Breit-Wigner line shape.

\begin{figure}[!t]
\begin{center}
\includegraphics[width=0.99\textwidth]{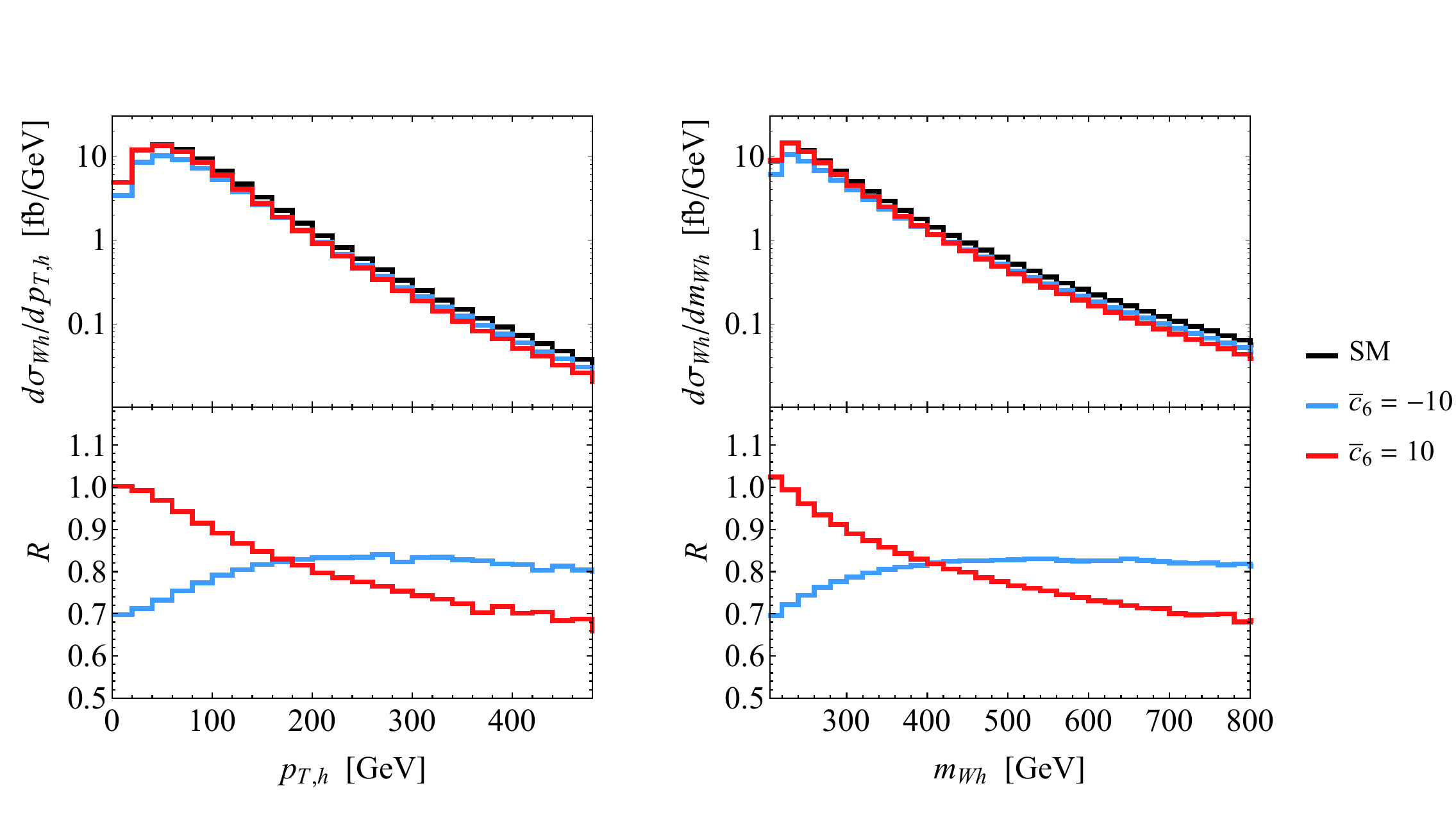}  
\vspace{-2mm}
\caption{\label{fig:whspectra}  Comparison of the $p_{T,h}$ (left) and $m_{Wh}$ (right) spectrum in $Wh$ production. The upper panels show the SM predictions (black) as well as the cases  $\bar c_6  = -10$ (blue)  and $\bar c_6  = 10$ (red). The ratios between the  case $\bar c_6  = -10$ and the SM (blue) and the case $\bar c_6  = 10$ and the SM (red) are displayed in the lower panels. All results correspond to $pp$ collisions at $\sqrt{s} = 13 \, {\rm TeV}$.}
\end{center}
\end{figure}

We begin our discussion with $pp \to Wh$. In Figure~\ref{fig:whspectra} the  distributions of the Higgs-boson transverse momentum ($p_{T,h}$) and the invariant mass of the $Wh$ system~($m_{Wh}$) are shown. The black curves in the panels represent the SM predictions, while the blue and red curves correspond to a new-physics scenario with $\bar c_6 = -10$ and $\bar c_6 = 10$, respectively. All results have been obtained at NNLO with the MC code described in Section~\ref{sec:Vh}. One sees that the shape of the displayed distributions provide sensitivity to the sign of $\bar c_6$. In the case of $\bar c_6 = -10$ the $p_{T,h}$ ($m_{Wh}$) spectrum increases relative to the SM distribution as a function of $p_{T,h}$ ($m_{Wh}$), approaching  a constant value  in the limit of large $p_{T,h}$ ($m_{Wh}$). For~$\bar c_6 = 10$ the ratio $R$ instead decreases with~$p_{T,h}$~($m_{Wh}$)  becoming again flat for $p_{T,h} \to \infty$ ($m_{Wh} \to \infty$). The behaviour of the distribution for large $p_{T,h}$ and  $m_{Wh}$ can be understood from the $\sqrt{s} \to \infty$ limit of~(\ref{eq:Vhcorrection}). In this limit  only the  Higgs wave function renormalisation contributes and the vertex correction~$\delta_V$  takes the simple form 
\beq \label{eq:deltaVlimit}
\lim_{\sqrt{s} \to \infty} \delta_V =  \frac{\lambda \hspace{0.25mm}  \bar c_6}{(4 \pi)^2} \left ( - 9 \hspace{0.25mm} m_h^2 \left ( \bar c_6 + 2 \right ) B_0^\prime \right ) = -1.5 \cdot 10^{-3} \; \bar c_6 \left ( \bar c_6 + 2 \right )  \,.
\eeq
It follows that for large transverse momenta (invariant masses) the deviation from 1 of  the ratio $R$ of the $p_{T,h}$~($m_{Wh}$) spectrum for $\bar c_6 \neq 0$ and $\bar c_6 = 0$,~i.e.~the SM distribution, is approximately given by (\ref{eq:deltaVlimit}). New-physics scenarios with $\bar c_6 < 0$ will hence lead to harder  $p_{T,h}$ and $m_{Wh}$  tails than cases with $\bar c_6 > 0$, while they predict softer spectra at low $p_{T,h}$ and $m_{Wh}$. These features are clearly visible in Figure~\ref{fig:whspectra} and are also present  in other kinematical observables such as the transverse momentum  $p_{T,W}$  of the $W$ boson. The shapes of all rapidity distributions in~$pp \to Wh$ production are in contrast largely insensitive to the sign of $\bar c_6$. Notice that our general arguments also apply  to the case of $pp \to Zh$, and as a result the distributions in the~$Zh$ channel resemble those found in $Wh$ production. We therefore do not show predictions for the various $Zh$ spectra. 

\begin{figure}[!t]
\begin{center}
\includegraphics[width=0.99\textwidth]{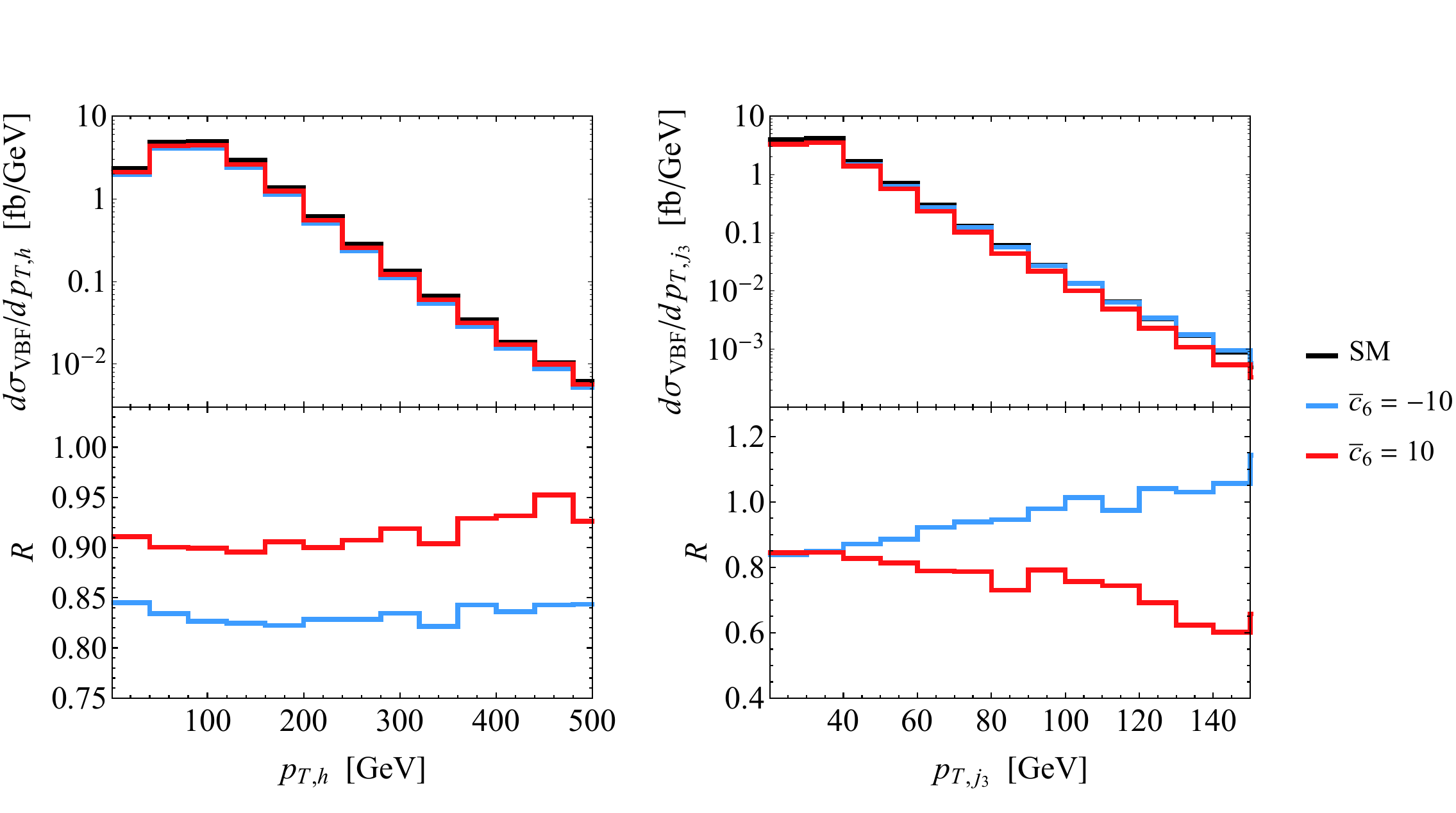}  
\vspace{-2mm}
\caption{\label{fig:VBFkinematics}  Comparison of the $p_{T,h}$ (left) and $p_{T,j_3}$ (right) spectrum in VBF Higgs production. The style and colour coding of the curves follows the one of  Figure~\ref{fig:whspectra}.}
\end{center}
\end{figure}

In Figure~\ref{fig:VBFkinematics} we present our results for two kinematic distributions in VBF Higgs production, namely the Higgs transverse momentum $p_{T,h}$ and the transverse momentum of the third jet~$p_{T,j_3}$. The spectra shown are obtained with the fully-differential NNLO  VBF code described in Section~\ref{sec:VBF} and correspond to the following  selection cuts. Events should have at least two jets with $p_{T,j} > 25 \, {\rm GeV}$, the two  jets with highest $p_{T,j}$ are required to have an absolute rapidity of $|y_j| < 4.5$, be separated by $\Delta y_{j_1, j_2} > 5$ in rapidity, have an invariant mass $m_{j_1, j_2} > 600 \, {\rm GeV}$ and be in opposite hemispheres (i.e.~$y_{j_1} y_{j_2} < 0$). In our analysis jets are defined using the anti-$k_t$ algorithm~\cite{Cacciari:2008gp}, as implemented in {\tt FastJet}~\cite{Cacciari:2011ma}, with radius parameter of $0.4$. As before we present results for the benchmark scenarios $\bar c_6 = -10$ (blue curves) and $\bar c_6 = 10$ (red curves) and compare them to the SM predictions (black curves). From the left panel in the figure we see that the shape of the $p_{T,h}$ distribution in VBF Higgs production is modified only mildly  by the presence of new physics in the~$h^3$~coupling, and  in consequence the ratio to the SM is almost constant in $p_{T,h}$. This feature can be explained by realising that even for large~$p_{T,h}$ one of the  squared momenta $Q_{1}^2$  or $Q_{2}^2$ that enters the form factors ${\cal F}_{1,2} (Q_1^2, Q_2^2)$~$\big($see~(\ref{eq:def})$\big)$ can be small. As a result for fixed $p_{T,h}$ a range of $Q_{1,2}^2$ values is probed and the constant ratio to the SM reflects this averaging. Similar averagings also take place  for instance for the transverse momentum of the first and second hardest jet, and hence the ratios $R$ corresponding to $p_{T,j_1}$ and $p_{T,j_2}$ turn out to be almost flat as well. On the contrary, when the third jet is hard both $Q_{1,2}^2$  tend to be hard. An increase in magnitude of both $Q_{1,2}^2$  gives rise to an approximately linear modification of the form factors ${\cal F}_{1,2} (Q_1^2, Q_2^2)$. This results in a linear shape in the ratio to the SM, as can be seen from the right panel in Figure~\ref{fig:VBFkinematics}. Still the effects are relatively small for the $p_{T,j_3}$ values accessible at the LHC, which will limit the discriminating power of shape analyses in the VBF Higgs production channel.

\subsection[Constraints on $\bar c_6$ from double-Higgs production]{Constraints on $\bm{\bar c_6}$ from double-Higgs production}

In the next subsection will derive the existing and possible future limits on the modifications of the trilinear Higgs-boson coupling that arise from $Vh$ and VBF Higgs production. All the numbers that we will present should be compared to the bounds that one can obtain by studying $pp \to hh$ production at the LHC. For definiteness we will assume throughout our numerical analysis that the modifications of the Wilson coefficient of the operator $O_6$ furnish the dominant contribution to the observable under consideration, and consequently neglect effects associated to other dimension-6 operators such as for instance $O_H$ --- see~(\ref{eq:operators}). 

The ATLAS collaboration has recently performed a search for  Higgs-boson pair production in the $2b 2\bar b$ final state using $13.3 \, {\rm fb}^{-1}$ of $\sqrt{s} = 13 \, {\rm TeV}$ data~\cite{ATLAS-CONF-2016-049}. From this measurement the cross section times branching ratio for non-resonant SM Higgs-boson pair production is constrained to be less than $330 \, {\rm fb}$, which is approximately  29 times above the SM expectation of  $ (\sigma_{2 b 2 \bar b}^{13 \, {\rm TeV}})_{\rm SM}  = (11.3^{+0.9}_{-1.0}) \, {\rm pb}$. By employing  {\tt HPAIR}~\cite{Grober:2015cwa,hpair}, we obtain 
\beq
\sigma_{2 b 2 \bar b}^{13 \, {\rm TeV}} = (\sigma_{2 b 2 \bar b}^{13 \, {\rm TeV}})_{\rm SM} \ \left ( 1 - 0.82 \hspace{0.25mm} \bar c_6 + 0.29  \hspace{0.25mm} \bar c_6^2 \right ) {\rm pb} \,.
\eeq
 From this formula we find that the ATLAS limit on the $p p \to 2 h \to 2 b 2 \bar b$ production cross section translates into the following 95\% confidence level (CL) bound 
\beq \label{eq:hhc6present}
\bar c_6 \in [ -9.5, 12.3 ]\,,
\eeq 
if theoretical uncertainties are taken into account. Note that (\ref{eq:hhc6present}) improves on the bound of $\bar c_6 \in  [-15.5, 18.1]$ that has been derived in~\cite{Gorbahn:2016uoy}  from  the ATLAS Run~I searches for $pp \to hh$~\cite{Aad:2014yja,Aad:2015uka,Aad:2015xja} by around $35\%$. It follows that the combination $\lambda c_3$ introduced in (\ref{eq:Ltriple}) can at present still deviate from the SM trilinear Higgs coupling $\lambda$ by a factor of roughly~11. 

The small rate, the mild dependence of the cross section on $\lambda$ and the difficulty of selecting signal from backgrounds make determinations of the trilinear Higgs coupling in $pp \to 2 h$ production challenging even at the HL-LHC. For instance  the ATLAS study of the $2b 2 \gamma$ final state~\cite{ATL-PHYS-PUB-2014-019} foresees a 95\% CL limit of 
\beq \label{eq:hhc6future}
\bar c_6 \in  [-2.3, 7.7] \,,
\eeq assuming~$3 \, {\rm ab}^{-1}$ of integrated luminosity.  Multivariate analyses (MVAs) and/or combinations of~$2b 2 \gamma$ with other decay channels such as $2\tau 2b$~\cite{ATL-PHYS-PUB-2015-046} or $2 b 2 \bar b$ may allow to  improve~(\ref{eq:hhc6future}), by how much precisely is however unclear at present. 

\subsection[Constraints on $\bar c_6$ from $Vh$ and VBF Higgs production]{Constraints on $\bm{\bar c_6}$  from $\bm{Vh}$ and VBF Higgs production}

Since only the product of the  production cross sections $\sigma_I$ and branching ratios~${\rm Br}^F$ of the Higgs boson can be extracted experimentally, it has become customary to define the signal strengths 
\beq \label{eq:muif}
\mu_I^F = \frac{\sigma_I}{(\sigma_I)_{\rm SM}} \, \frac{{\rm Br}^F}{({\rm Br}^F)_{\rm SM}}\,,
\eeq
which characterise the Higgs boson yields in a specific production and decay channel relative to the SM expectations. The formalisms  of signal strengths can then be used to test the compatibility of the LHC measurements with the SM and to interpret the Higgs data in the context of BSM searches. 

To obtain the current constraints on $\bar c_6$ we use the LHC Run I combination of the ATLAS and CMS measurements of the Higgs boson production and decay rates~\cite{ATLASCMS}.  In the case of the vector boson mediated production processes the relevant $\mu_I^F$ parameters read
\beq \label{eq:signalstrengths16}
\begin{split}
& \hspace{1.75cm} \mu_{V}^{b\bar b} = 0.65^{+0.30}_{-0.29} \,, \qquad 
\mu_{V}^{WW} = 1.38^{+0.41}_{-0.37} \,,  \\[2mm]
& \mu_{V}^{\tau^+ \tau^-} = 1.12^{+0.37}_{-0.35} \,, \qquad 
\mu_{V}^{ZZ} = 0.48^{+1.37}_{-0.91} \,, \qquad 
\mu_{V}^{\gamma \gamma} =1.05^{+0.44}_{-0.41} \,, 
\end{split}
\eeq
where the subscript $V$ indicates that the above numbers correspond to a combination of the $Vh$ and ${\rm VBF}$ channels. These numbers have been obtained from a 10-parameter fit  to each of the five decay channels and can be found in the upper part of Table 13 of~\cite{ATLASCMS}. The quoted uncertainties take into account the experimental uncertainty in the measurement of~$\mu_I^F$ as well as the SM theory error associated to each particular channel.  In the following we will employ this framework to set limits on the Wilson coefficient $\bar c_6$. 

Using our predictions for $\sigma_I$ and ${\rm Br}^F$ presented in Sections~\ref{sec:sigmaI} and \ref{sec:BrF} we then can calculate the signal strengths $\mu_I^F$ and compare them to experiment. Including the errors quoted in (\ref{eq:signalstrengths16}) but neglecting theoretical uncertainties associated to  missing $\lambda$ terms, we obtain  the limit 
\beq \label{eq:c6current}
\bar c_6 \in  [-13.6, 16.9] \,, \quad {\text{(LHC Run I)}} \,,
\eeq 
by performing a $\chi^2$ fit with $\Delta \chi^2 = 3.84$ which corresponds to a 95\% CL for a Gaussian distribution. This constraint is somewhat weaker than both the bound (\ref{eq:hhc6present})   as well as the limit of $\bar c_6 \in [-11.9, 10.3]$ that follows from a combination of the~$gg \to h$ and $h \to \gamma \gamma$ channels~\cite{Gorbahn:2016uoy,inprep}. Notice that~our bound (\ref{eq:c6current}) compares well with the current limits on the modifications of the trilinear Higgs coupling reported in \cite{Degrassi:2016wml}.

The experimental prospects for measuring the Higgs boson signal strengths (\ref{eq:muif}) in the vector boson mediated production modes at future LHC runs has been studied by both the ATLAS and CMS collaborations~\cite{ATL-PHYS-PUB-2014-011,ATL-PHYS-PUB-2014-012,ATL-PHYS-PUB-2014-016,ATL-PHYS-PUB-2014-018,ATL-PHYS-PUB-2016-008,CMS:2013xfa}. To estimate the sensitivity on $\bar c_6$ that can be reached at the HL-LHC with $3 \, {\rm ab}^{-1}$ of data, we study two benchmark scenarios based on the results  reported  in the fourth and fifth column of Table 1 of~\cite{ATL-PHYS-PUB-2014-016}.\footnote{The inclusion of further channels such as  for instance $pp \to Vh \, (h \to \tau^+ \tau^-)$~\cite{Boddy:2012nt} or technical developments like extended jet tracking \cite{CERN-LHCC-2015-020} are expected to result in an improved precision on the signals strengths~$\mu_I^F$. In order to obtain a conservative future limit on the Wilson coefficient $\bar c_6$ we do not consider  such improvements.} Our first scenario includes the current theory  uncertainties and reads 
\bea \label{eq:deltasignalstrengthsf1} 
\begin{split}
&  \hspace{3.5cm} \Delta \mu_{Wh}^{b\bar b} = \pm 37\% \,, \quad
\Delta \mu_{Wh}^{\gamma \gamma} = \pm 19\%  \,, \\[2mm]
&  \hspace{2.1cm} \Delta \mu_{Zh}^{b\bar b} = \pm 14\% \,, \quad
\Delta \mu_{Zh}^{\gamma \gamma} = \pm 28\%  \,, \quad 
\Delta \mu_{Vh}^{ZZ} = \pm 13\%\,, \\[2mm]
&  \Delta \mu_{\rm VBF}^{WW} = \pm 15\%\,, \quad
\Delta \mu_{\rm VBF}^{\tau^+ \tau^-} = \pm 19\%\,, \quad
\Delta \mu_{\rm VBF}^{ZZ} = \pm 21\%\,, \quad
\Delta \mu_{\rm VBF}^{\gamma \gamma} = \pm 22\%  \,,
\end{split}
\eea
whereas in the second benchmark scenario theoretical errors are not taken into account. The corresponding relative uncertainties are 
\bea \label{eq:deltasignalstrengthsf2}
\begin{split}
&  \hspace{3.5cm} \Delta \mu_{Wh}^{b\bar b} = \pm 36\% \,, \quad
\Delta \mu_{Wh}^{\gamma \gamma} = \pm 17\%  \,, \\[2mm]
&  \hspace{2.1cm} \Delta \mu_{Zh}^{b\bar b} = \pm 13\% \,, \quad
\Delta \mu_{Zh}^{\gamma \gamma} = \pm 27\%  \,, \quad 
\Delta \mu_{Vh}^{ZZ} = \pm 12\%\,, \\[2mm]
&  \Delta \mu_{\rm VBF}^{WW} = \pm 9\%\,, \quad
\Delta \mu_{\rm VBF}^{\tau^+ \tau^-} = \pm 15\%\,, \quad
\Delta \mu_{\rm VBF}^{ZZ} = \pm 16\%\,, \quad
\Delta \mu_{\rm VBF}^{\gamma \gamma} = \pm 15\%  \,.
\end{split}
\eea
Notice that compared to the CMS projections \cite{ATL-PHYS-PUB-2016-008} our~HL-LHC benchmark uncertainties~$ \Delta \mu_{I}^{F} $ are comparable but in all cases slightly larger, irrespectively of whether or not theory errors are included in the final numbers.  

Assuming that the central values of the future HL-LHC measurements coincide in every channel with the predictions of the SM, we obtain the following 95\% CL limit on the Wilson coefficient of $O_6$ from our $\chi^2$ fit
\beq \label{eq:c6future1}
\bar c_6 \in  [-7.0, 10.9] \,, \quad {\text{(HL-LHC, all uncertainties)}} \,, 
\eeq 
when all uncertainties are included. If theoretical errors are neglected, we instead find
\beq \label{eq:c6future2}
\bar c_6 \in  [-6.2, 9.6] \,, \quad {\text{(HL-LHC, no theory uncertainty)}} \,.
\eeq 
These limits improve on the current constraint (\ref{eq:c6current})  by a factor of   around 1.7 to 2, depending on how theory errors are treated. They should be compared to the determination~(\ref{eq:hhc6future}) of~$\bar c_6$ in double-Higgs production.  We see that with the full HL-LHC data set the indirect determination of $\bar c_6$ through measurements of $pp \to Vh$ and $pp \to jj h$ should allow to test shifts in the trilinear Higgs coupling that are at the same level than the more direct extraction via $pp \to hh$. A comparison of (\ref{eq:c6future1}) and (\ref{eq:c6future2}) also shows that theoretical uncertainties  are not a limiting factor for the extraction of $\bar c_6$ through measurements of $Vh$ and VBF Higgs production.

We finally add that future LHC combinations of  the cross section measurements of $pp \to Vh$ and $pp \to jj h$  with those of $gg \to h$~\cite{Gorbahn:2016uoy,Degrassi:2016wml} and $pp \to t \bar t h$ \cite{Degrassi:2016wml} are expected to  further strengthen the indirect constraints on the Wilson coefficient of the operator $O_6$. Differential information from single Higgs production and/or decays may  also be used to improve the sensitivity on $\bar c_6$. Making the latter statement more precise would require a  MVA of the prospects to measure $Vh$ and VBF Higgs distributions in the HL-LHC environment building on the results presented in~Section~\ref{sec:distributions}. Such a study  is however beyond the scope of this article.

\section{Conclusions}
\label{sec:conclusions}

The main goal of this work was to  constrain possible deviations in the $h^3$ coupling using measurements of $Vh$ and VBF  Higgs production in $pp$ collisions. In order to keep the entire discussion model independent, we have adopted the SMEFT framework, in which the effects of new heavy particles are encoded in the Wilson coefficients of higher-dimensional operators. Within the SMEFT, we have calculated the ${\cal O} (\lambda)$ corrections to  the~$pp  \to Vh$ and~$pp \to jj h$ amplitudes that arise from insertions of the operator $O_6 = -\lambda \left (H^\dagger H \right)^3$ into 1-loop Feynman diagrams. We have supplemented this calculation by a computation of the ${\cal O} (\lambda)$ corrections to the partial decay widths of the Higgs boson in  $h \to f \bar f$, $h \to VV$, $h \to gg$ and $h \to \gamma \gamma$. By combining both calculations we are able to derive the full ${\cal O} (\lambda)$ corrections to all  phenomenological relevant vector boson mediated Higgs signal strengths. 

To obtain accurate predictions for the $pp  \to Vh$ and~$pp \to jj h$  our MC simulations include~QCD corrections up to NNLO. We have studied the impact of a modified $h^3$ vertex on the inclusive  cross sections and the most important kinematic distributions in $Vh$ and~VBF~Higgs production. The dependencies of the inclusive production cross sections on~$\bar c_6$ turn out to be process dependent and slightly stronger  in the $Vh$ channels than in VBF Higgs production. Since the ${\cal O} (\lambda)$ corrections to the $VVh$ vertex depend in a non-trivial way on the external 4-momenta, the $\bar c_6$ dependence is also sensitive to the kinematic configurations of the final state under consideration.  Our study  of kinematic distributions in $pp  \to Vh$ and~$pp \to jj h$ shows that the shapes of the transverse momentum or invariant mass spectra in these channels are sensitive to both the  size and sign of $\bar c_6$. However a more detailed analysis than the one performed in our article is required to determine to which extent differential information in $Vh$ and VBF Higgs production can be used to improve the constraints on~$\bar c_6$ that can be derived using inclusive rates. We plan to return to this question in future work.

Under the assumption that~$\bar c_6$ is the only Wilson coefficient that obtains a non-zero correction in the SMEFT, we have then studied the sensitivity of present and future LHC measurements of  $Vh$ and VBF  Higgs production to a modified $h^3$ interaction. We have first demonstrated that the constraint on $\bar c_6$ that follows from a combination of the~LHC~Run~I measurements of signal strengths in $Vh$ and VBF Higgs production are slightly more stringent than the  limit obtained from double-Higgs production using Run I data.  In the case of the HL-LHC with $3 \, {\rm ab}^{-1}$ of integrated luminosity, we have furthermore found that it should be possible to improve the present bound by a factor of at least  1.7. As a result indirect  determinations of  $|\bar c_6|  \lesssim 9$ based on $Vh$ and VBF  Higgs production data alone should be possible. This conservative limit is not significantly weaker than the bound obtained by the~ATLAS sensitivity study~\cite{ATL-PHYS-PUB-2014-019} from double-Higgs production at the HL-LHC. 

Further improvements of the constraints on the trilinear Higgs coupling are possible by combining  the signal strength measurements in $pp \to Vh$ and $pp \to jj h$  with those in $gg \to h$~\cite{Gorbahn:2016uoy,Degrassi:2016wml} and $pp \to t \bar t h$ \cite{Degrassi:2016wml}. The  indirect probes of the trilinear Higgs coupling studied here and  in \cite{Gorbahn:2016uoy,Degrassi:2016wml} hence provide information that is complementary  to the direct  determinations of $\lambda$ through $pp \to hh$ production. Since the indirect and direct tests constrain different linear combinations of effective operators in the SMEFT, we believe that it is crucial to combine  all available information on the $h^3$ coupling in the form of a global fit to fully exploit the potential of the HL-LHC. We look forward to further theoretical but also experimental investigations in this direction.  

\acknowledgments 
We thank Alexander~Karlberg for providing optimised routines  for the VBF $h + 3 \, {\rm jets}$ phase space and matrix elements.  We are grateful to Chris Hays for  helpful discussions and correspondence concerning the experimental precision that measurements of $Vh$ and VBF Higgs production may reach at the~HL-LHC and to Matthew McCullough for valuable feedback concerning  his publication \cite{McCullough:2013rea}.  WB,~UH and GZ have been partially supported by the ERC grant 614577 ``HICCUP --- High Impact Cross Section Calculations for Ultimate Precision'', while the research of MG has been funded by the STFC consolidated grant ST/L000431/1. UH~would like to thank the CERN Theoretical Physics Department for continued hospitality and support. UH~and~GZ are  finally grateful to the~MITP in Mainz for its hospitality and its partial support during the initial phase  of this work.

\end{document}